\documentclass[11pt]{article}
\usepackage{geometry}
\usepackage{amsfonts}
\usepackage{amsmath}
\usepackage{amssymb}
\usepackage{mathptmx}
\usepackage{color}
\usepackage{epsfig}
\usepackage{graphicx}
\usepackage{hyperref}
\usepackage{cancel}
\usepackage{textcomp}
\usepackage{bbold}
\usepackage{bm}
\usepackage{times}
\usepackage{slashed}
\usepackage{booktabs}
\usepackage{verbatim}
\usepackage{latexsym}
\usepackage{extarrows}
\usepackage{multirow}
\usepackage[dvipsnames]{xcolor}

\newcommand{\T}{{\rm T}}

\newcommand{\nsmeft}{\nu\textrm{SMEFT}}

\begin{document}

\baselineskip=16pt

\pagenumbering{arabic}

\vspace{1.0cm}

\begin{center}
{\Large\sf Operators up to dimension seven in standard model effective field theory extended with sterile neutrinos}
\\[10pt]
\vspace{.5 cm}

{Yi Liao~$^{a,b,c}$\footnote{liaoy@nankai.edu.cn} and Xiao-Dong Ma~$^{a}$\footnote{maxid@mail.nankai.edu.cn}}

{
$^a$~School of Physics, Nankai University, Tianjin 300071, China
\\
$^b$ CAS Key Laboratory of Theoretical Physics, Institute of Theoretical Physics,
Chinese Academy of Sciences, Beijing 100190, China
\\
$^c$ Center for High Energy Physics, Peking University, Beijing 100871, China}

\vspace{2.0ex}

{\bf Abstract}

\end{center}

We revisit the effective field theory of the standard model that is extended with sterile neutrinos, $N$. We examine the basis of complete and independent effective operators involving $N$ up to mass dimension seven (dim-7). By employing equations of motion, integration by parts, and Fierz and group identities, we construct relations among operators that were considered independent in the previous literature, and find seven redundant operators at dim-6, sixteen redundant operators and two new operators at dim-7. The correct numbers of operators involving $N$ are, without counting Hermitian conjugates, $16~(L\cap B)+1~(\slashed{L}\cap B)+2~(\slashed{L}\cap\slashed{B})$ at dim-6, and $47~(\slashed{L}\cap B)+5~(\slashed{L}\cap\slashed{B})$ at dim-7. Here $L/B~(\slashed L/\slashed B)$ stands for lepton/baryon number conservation (violation). We verify our counting by the Hilbert series approach for $n_f$ generations of the standard model fermions and sterile neutrinos. When operators involving different flavors of fermions are counted separately and their Hermitian conjugates are included, we find there are $29~(1614)$ and $80~(4206)$ operators involving sterile neutrinos at dim-6 and dim-7 respectively for $n_f=1~(3)$.

\newpage

\section{Introduction}
\label{sec1}

Recent years have witnessed great progress in the field of standard model (SM) effective field theory (EFT). The basic idea is that assuming there are no new particles below the electroweak scale $\Lambda_\textrm{EW}$ the effects from new physics above $\Lambda_\textrm{EW}$ can be incorporated into a tower of higher dimensional operators. These operators are built out of the SM fields, respect the SM gauge symmetry $SU(3)_C\times SU(2)_L\times U(1)_Y$, and generically get more and more suppressed by new physics scale as their mass dimension increases. The bases for complete and independent operators up to dimension seven (dim-7) have now been established after years of efforts~\cite{Weinberg:1979sa,Buchmuller:1985jz,Grzadkowski:2003tf,Fox:2007in,AguilarSaavedra:2009mx,
Grzadkowski:2010es,Weinberg:1980bf,Weldon:1980gi,Lehman:2014jma,Liao:2016hru}. This conventional task of manually sorting out operators and checking their completeness and independence has recently got a strong support from the so-called Hilbert series in invariant theory, and effective operators of even higher mass dimension have been studied, see Refs.~\cite{Lehman:2015via,Henning:2015daa,Lehman:2015coa,Henning:2015alf} and references cited therein. Although the Hilbert series approach does not decide on manifest gauge and Lorentz structures for operators, it counts the number of complete and independent operators that can be formed for each combination of fields and derivatives and for any generations of fermions. More importantly from the practical point of view, the counting can be programmed~\cite{Henning:2015alf}. This offers a very useful handle to check the results obtained in the conventional approach that removes redundancy of operators by judiciously employing integration by parts (IBP), equations of motion (EoM), Fierz and group identities and that could not automatically establish the completeness of operators.

On the other hand, the existence of neutrino mass and mixing and dark matter provides definite evidence for physics beyond SM. The easiest way to incorporate both would be to introduce sterile neutrinos. Noting that relatively light sterile neutrinos have not yet been excluded for general parameters, it is possible that they may have a mass below $\Lambda_\textrm{EW}$. In that case it makes sense to consider a low energy effective field theory that includes SM particles and sterile neutrinos in the same setting, named $\nsmeft$ for short. The construction of additional effective operators in $\nsmeft$ on top of those in standard model effective field theory (SMEFT) has appeared in the literature, see Ref.~\cite{Aparici:2009fh} for dim-5, Ref.~\cite{delAguila:2008ir} for dim-6, and Ref.~\cite{Bhattacharya:2015vja} for dim-7 operators respectively. Some of their astrophysical and collider implications have received recent attention~\cite{Aparici:2009fh,delAguila:2008ir,Bhattacharya:2015vja,Aparici:2009oua,ATLAS:2012ak,
Peressutti:2014lka,Duarte:2014zea,Duarte:2015iba,Ballett:2016opr}. In this context however, we would like to recall that it is important to work with a basis of complete and independent operators because operators related by, e.g., EoM, will contribute the same to the $S$ matrix according to equivalence theorem~\cite{Kamefuchi:1961sb,Salam:1971sp,Kallosh:1972ap,Bergere:1975tr,Arzt:1993gz}.

In this work, we revisit higher dimensional operators in $\nsmeft$ that involve at least one sterile neutrino field. These operators are to be added to those in SMEFT to form a basis of complete and independent operators at each mass dimension in $\nsmeft$. Our study shows that there are redundant operators at both dim-6~\cite{delAguila:2008ir} and dim-7~\cite{Bhattacharya:2015vja} and that there are missing dim-7 operators~\cite{Bhattacharya:2015vja}. Specifically, we found seven dim-6 redundant operators, sixteen dim-7 redundant operators and two new dim-7 operators. The correct number of operators involving sterile neutrinos in $\nsmeft$ is thus, $16~(L\cap B)+1~(\slashed{L}\cap B)+2~(\slashed{L}\cap\slashed{B})$ at dim-6 and $47~(\slashed{L}\cap B)+5~(\slashed{L}\cap\slashed{B})$ at dim-7, without counting their Hermitian conjugates. Here $L/B~(\slashed L/\slashed B)$ stands for lepton/baryon number conservation (violation). We establish this result by systematically sorting out possible operators at each mass dimension and removing redundant ones with extensive applications of IBP, EoM, and in particular Fierz and group identities. We further confirm our counting for general $n_f$ generations of fermions with a slight modification to the Mathematica code in Ref.~\cite{Henning:2015alf} in the Hilbert series approach. For instance, when operators differing in fermion flavors are also separately counted and Hermitian conjugates of non-Hermitian operators are also included, there are a total number of 29 (1614) operators at dim-6 and 80 (4206) operators at dim-7 for $n_f=1~(3)$ that involve sterile neutrino fields. Our basis of complete and independent operators up to dim-7 provides an appropriate starting point for consistent phenomenological analysis, to which we hope to come back in the future.

The paper is organized as follows. Section~\ref{sec2} deals with dim-6 operators, where we build several relations to be used for the demonstration of redundancy of operators in Ref.~\cite{delAguila:2008ir}. In Sec.~\ref{sec3}, we perform a systematic search for dim-7 operators in terms of the number of sterile neutrino fields involved, and make a detailed comparison with the operator basis in Ref.~\cite{Bhattacharya:2015vja}. We conclude briefly in the last Sec.~\ref{sec4}.

\section{Dimension six operators involving sterile neutrinos}
\label{sec2}

We start with some notational preparations. We introduce for simplicity one sterile neutrino $N$ per generation of the SM fermions ($Q,~u,~d,~L,~e$) although two would be enough to generate two light neutrino masses at the tree level. Without loss of generality, we assume $N$ to be right-handed. The renormalizable Lagrangian is,
\begin{eqnarray}
\label{sml}
\nonumber
\mathcal{L}_4&=&
-\frac{1}{4}G^A_{\mu\nu}G^{A\mu\nu}-\frac{1}{4}W^I_{\mu\nu}W^{I\mu\nu} -\frac{1}{4}B_{\mu\nu}B^{\mu\nu}+(D_\mu H)^\dagger(D^\mu H)
-\lambda\left(H^\dagger H-\frac{1}{2}v^2\right)^2
\\
&&+\sum_{\Psi=Q, L, u, d, e, N}\bar{\Psi}i \slashed{D}\Psi
-\left[\frac{1}{2}\left(NCM_NN\right)+\bar{L}Y_NN\tilde{H}+\bar{L}Y_e e H
+\bar{Q}Y_u u \tilde{H}+\bar{Q}Y_d d H +\mbox{h.c.}\right],
\end{eqnarray}
where $H$ is the Higgs doublet with the vacuum expectation value $v$ and $\tilde H_i=\epsilon_{ij}H^*_j$, and $G^A_{\mu\nu},~W^I_{\mu\nu},~B_{\mu\nu}$ are the gauge field strength tensors. $M_N$ is the symmetric mass matrix of $N$ and $Y_{u,d,e,N}$ are the Yukawa coupling matrices. $C$ is the charge-conjugation matrix, and $D_\mu$ is the gauge covariant derivative appropriate for each field which for $N$ is the ordinary partial derivative $\partial_\mu$.

Considering the extended SM as a low energy effective field theory, i.e., $\nsmeft$, the above Lagrangian will be augmented by a tower of higher dimensional operators,
\begin{equation}
\mathcal{L}_{\nsmeft}=\mathcal{L}_4+\mathcal{L}_5+\mathcal{L}_6+\mathcal{L}_7+\cdots.
\end{equation}
The final list of operators without involving a sterile neutrino field $N$, i.e., within SMEFT, is explicitly given in Ref.~\cite{Weinberg:1979sa} for dim-5, Ref.~\cite{Grzadkowski:2010es} for dim-6, and Ref.~\cite{Liao:2016hru} for dim-7 operators, respectively, while Refs.~\cite{Lehman:2015coa,Henning:2015alf} studied even higher dimensional operators in the Hilbert series approach. In this work, we focus on the additional operators up to dim-7 that involve at least one factor of $N$. Such dim-5 operators were easily found to be, $(NCN)(H^\dagger H)$ and $(NC\sigma_{\mu\nu}N)B^{\mu\nu}$ (plus their Hermitian conjugates)~\cite{Aparici:2009fh}, while dim-6 and dim-7 ones were studied previously in Refs.~\cite{delAguila:2008ir} and \cite{Bhattacharya:2015vja} respectively. But as we will show in this and next section, those dim-6 and dim-7 operators are redundant and in addition the dim-7 operators are incomplete.

\begin{table}
\centering
\begin{tabular}{|c|c|c|c|c|c|}
\hline
 \multicolumn{2}{|c|}{$\psi^2H^3$} & \multicolumn{2}{|c|}{$\psi^2H^2D$} & \multicolumn{2}{|c|}{$\psi^2HX(+\mbox{h.c.})$}
\\
\hline
$\mathcal{O}_{LNH}(+\mbox{h.c.})$ & $(\bar{L}N)\tilde{H}(H^\dagger H)$ &
$\mathcal{O}_{HN}$ &  $(\bar{N}\gamma^\mu N)(H^\dagger i \overleftrightarrow{D_\mu} H)$&
$\mathcal{O}_{NB}$ & $(\bar{L}\sigma_{\mu\nu}N)\tilde{H}B^{\mu\nu}$
\\
& &
$\mathcal{O}_{HNe}(+\mbox{h.c.})$ & $(\bar{N}\gamma^\mu e)({\tilde{H}}^\dagger i D_\mu H)$ &
$\mathcal{O}_{NW}$ &$(\bar{L}\sigma_{\mu\nu}N)\tau^I\tilde{H}W^{I\mu\nu}$
\\
\hline
\multicolumn{2}{|c|}{$(\bar{R}R)(\bar{R}R)$}  &   \multicolumn{2}{|c|}{$(\bar{L}L)(\bar{R}R)$} &   \multicolumn{2}{|c|}{$(\bar{L}R)(\bar{L}R)(+\mbox{h.c.})$}
\\
\hline
$\mathcal{O}_{NN}$ & $(\bar{N}\gamma^\mu N)(\bar{N}\gamma_\mu N)$ &
$\mathcal{O}_{LN}$ & $(\bar{L}\gamma^\mu L)(\bar{N}\gamma_\mu N)$ &
$\mathcal{O}_{LNLe}$ & $(\bar{L}N)\epsilon(\bar{L}e)$
\\
$\mathcal{O}_{eN}$ & $(\bar{e}\gamma^\mu e)(\bar{N}\gamma_\mu N)$ &
$\mathcal{O}_{QN}$ & $(\bar{Q}\gamma^\mu Q)(\bar{N}\gamma_\mu N)$ &
$\mathcal{O}_{LNQd}$ & $(\bar{L}N)\epsilon(\bar{Q}d))$
\\
$\mathcal{O}_{uN}$ & $(\bar{u}\gamma^\mu u)(\bar{N}\gamma_\mu N)$ &
& &
$\mathcal{O}_{LdQN}$ & $(\bar{L}d)\epsilon(\bar{Q}N)$
\\
$\mathcal{O}_{dN}$ & $(\bar{d}\gamma^\mu d)(\bar{N}\gamma_\mu N)$&
& &
&
\\
$\mathcal{O}_{duNe}(+\mbox{h.c.})$ & $ (\bar{d}\gamma^\mu u)(\bar{N}\gamma_\mu e)$&
& &
&\\
\hline
\multicolumn{2}{|c|}{$(\bar{L}R)(\bar{R}L)$}  &
\multicolumn{2}{|c|}{$(\slashed{L}\cap B)(+\mbox{h.c.})$}  &   \multicolumn{2}{|c|}{$(\slashed{L}\cap\slashed{B})(+\mbox{h.c.})$}
\\ \hline
$\mathcal{O}_{QuNL}(+\mbox{h.c.})$ & $(\bar{Q}u)(\bar{N}L)$ &
$\mathcal{O}_{NNNN}$ & $(NCN)(NCN)$ &
 $\mathcal{O}_{QQdN}$ & $\epsilon_{ij}\epsilon_{\alpha\beta\sigma}(Q^i_{\alpha}CQ^j_{\beta})(d_{\sigma}CN)$
\\
& &
& &
$\mathcal{O}_{uddN}$ & $\epsilon_{\alpha\beta\sigma}(u_{\alpha}Cd_{\beta})(d_{\sigma}CN)$
\\
\hline
\multicolumn{6}{c}{Redundant operators}
\\
\hline
$\mathcal{O}_{LNNL}$ & $(\bar{L}N)(\bar{N}L)$&
$\mathcal{O}_{QNNQ}$ & $(\bar{Q}N)(\bar{N}Q)$ &
$\mathcal{O}^\prime_{NN}$ & $(\bar{N}N^C)(NCN)$
\\
$\mathcal{O}_{QNdQ}(+\mbox{h.c.})$ & $(\bar{Q}N^C)(\bar{d}Q^C)$ &
$\mathcal{O}_{uNd}(+\mbox{h.c.})$ & $\epsilon_{\alpha\beta\sigma}(\bar{u}_{\alpha}N^C)(\bar{d}_{\beta}d^C_{\sigma})$ &
&
\\
$\mathcal{O}_{DN}(+\mbox{h.c.})$ & $(\bar{L}D_\mu N)D^\mu\tilde{H}$ &
$\mathcal{O}_{\bar{D}N}(+\mbox{h.c.})$ & $(\bar{L}\overleftarrow{D}_\mu N)D^\mu\tilde{H}$
&  &
\\
\hline
\end{tabular}
\caption{The 19 complete and independent dim-6 operators involving $N$ named similarly to Refs.~\cite{Grzadkowski:2003tf,Liao:2016hru} are shown in the upper part of the table while the 7 redundant ones named as in~\cite{delAguila:2008ir} are in the lower part. The notation $(+\mbox{h.c.})$ indicates the Hermitian conjugates of relevant operators, and $\alpha,~\beta,~\sigma$ ($i,~j$) are $SU(3)_C$ ($SU(2)_L$) indices.}
\label{tab1}
\end{table}

The work in~\cite{delAguila:2008ir} made a systematic study of dim-6 operators and found 26 operators (without counting Hermitian conjugates of non-Hermitian ones). These operators are listed in Table~\ref{tab1} in two categories according to our result, i.e., the complete and independent 19 operators vs 7 redundant ones. Here we follow as closely as possible the notations for fields and operators in Refs.~\cite{Grzadkowski:2003tf,Liao:2016hru}. That there are only two operators violating baryon number by one unit is also consistent with Ref.~\cite{Alonso:2014zka}. To prove our claim, we will need the following Fierz identities that were derived in~\cite{Liao:2016hru} based on Refs.~\cite{Liao:2012uj,Nieves:2003in}:
\begin{eqnarray}
\label{fierz1}
( \overline{\Psi_{1L}} \gamma^\mu \Psi_{2L})( \overline{\Psi_{3L}}  \gamma_\mu \Psi_{4L})
&=&2(\overline{\Psi_{1L}} \Psi^C_{3L})(\overline{\Psi^C_{4L}}\Psi_{2L}),
\\
\label{fierz2}
( \overline{\Psi_{1L}} \gamma^\mu \Psi_{2L})( \overline{\Psi_{3R}}  \gamma_\mu \Psi_{4R})
&=&-2(\overline{\Psi_{1L}} \Psi_{4R})(\overline{\Psi_{3R}}\Psi_{2L}),
\\
\label{fierz3}
( \overline{\Psi_{1R}}  \Psi_{2L})( \overline{\Psi^C_{3L}}  \Psi_{4L})
&=&-( \overline{\Psi_{1R}}\Psi_{3L})( \overline{\Psi^C_{4L}}  \Psi_{2L})-( \overline{\Psi_{1R}}  \Psi_{4L})( \overline{\Psi^C_{3L}} \Psi_{2L}),
\\
\label{fierz4}
( \overline{\Psi_{1R}} \gamma^\mu \Psi_{2R})( \overline{\Psi^C_{3L}}  \Psi_{4L})
&=&( \overline{\Psi_{1R}}  \Psi_{3L})( \overline{\Psi^C_{2R}}  \gamma_\mu \Psi_{4L})+( \overline{\Psi_{1R}}  \Psi_{4L})( \overline{\Psi^C_{2R}}  \gamma_\mu \Psi_{3L}),
\end{eqnarray}
where $\Psi^C_L=C\overline{\Psi_L}^T$ and anticommutativity of fermion fields has been considered. The identities also hold true on chirality flip $\Psi_L\leftrightarrow\Psi_R$.

We first reduce the five redundant operators not involving a derivative as a direct consequence of the above Fierz identities. We attach flavor indices $p,~r,~v,~w$ to fields and operators to show better the shift of flavors:
\begin{eqnarray}
\nonumber
\mathcal{O}^{prst}_{LNNL}&\overset{(\ref{fierz2})}{=}&-\frac{1}{2}\mathcal{O}^{ptsr}_{LN},
\\
\nonumber
\mathcal{O}^{prst}_{QNNQ}&\overset{(\ref{fierz2})}{=}&-\frac{1}{2}\mathcal{O}^{ptsr}_{QN},
\\
\nonumber
\mathcal{O}^{\prime prst}_{NN}&\overset{(\ref{fierz1})}{=}&\frac{1}{2}\mathcal{O}^{ptrs}_{NN},
\\
(\mathcal{O}^{prst}_{uNd})^\dagger&\overset{(\ref{fierz3})}{=}&
\mathcal{O}^{ptsr}_{uddN}-\mathcal{O}^{pstr}_{uddN},
\end{eqnarray}
where the operators on the right-hand side are among the 19 operators in Table~\ref{tab1}, while $\mathcal{O}^{ prst}_{QNdQ}=0$ trivially because of chirality mismatch. To prove the redundancy of the operators involving derivatives, we require the well-known relations
\begin{eqnarray}
\label{dirac1}
2g_{\mu\nu}&=&\{\gamma_\mu,\gamma_\nu\},
\\
\label{dirac2}
\gamma_\mu\gamma_\nu&=&g_{\mu\nu}-i\sigma_{\mu\nu},
\end{eqnarray}
as well as IBP and EoM, so that we can transform them in steps:
\begin{eqnarray}
\mathcal{O}_{\bar{D}N}&\overset{\mbox{IBP}}{=}&-\mathcal{O}_{DN}-(\bar{L} N)D^2\tilde{H}
\nonumber\\
\label{dn1}
&\overset{\mbox{EoM}}{=}&-\mathcal{O}_{DN}+\dots,
\end{eqnarray}
and
\begin{eqnarray}
\nonumber
2\mathcal{O}_{DN}
&\overset{\mbox{(\ref{dirac1})}}{=}&
(\bar{L}\slashed{D} \gamma_\mu N) D^\mu\tilde{H}+(\bar{L}\gamma_\mu \slashed{D}N) D^\mu\tilde{H}
\\
\nonumber
&\overset{\mbox{IBP}}{=}&-(\bar{L}  \gamma_\mu \gamma_\nu N) D^\mu D^\nu\tilde{H}-(\bar{L}\overleftarrow{\slashed{D}} \gamma_\mu N) D^\mu\tilde{H}+(\bar{L}\gamma_\mu \slashed{D}N) D^\mu\tilde{H}
\\
\nonumber
&\overset{\mbox{(\ref{dirac2})}}{=}&-(\bar{L} N) D^2\tilde{H}+\frac{i}{2}(\bar{L}\sigma_{\mu\nu} N) [D^\mu, D^\nu]\tilde{H}-(\bar{L}\overleftarrow{\slashed{D}} \gamma_\mu N) D^\mu\tilde{H}+(\bar{L}\gamma_\mu \slashed{D}N) D^\mu\tilde{H}
\\
\label{dn2}
&\overset{\mbox{EoM}}{=}&\dots,
\end{eqnarray}
where the dots stand for the operators already covered in our basis for $\nsmeft$. This establishes our claim.

\begin{table}
\small
\centering
\begin{tabular}{|c|c|c|c|c|c|}
\hline
Class & Operator &  Symmetry relation &$n_f$&  $n_f=1$ &$n_f=3$
\\
\hline
$\psi^2H^3+\mbox{h.c.}$ &$\mathcal{O}_{LNH}$ & $\times$ & $2n_f^2$ & $2$ &$18$
\\
\hline
$\psi^2H^2D$&$\mathcal{O}_{HN}$  & $\times$ & $n_f^2$ & $1$ & $9$
\\
& $\mathcal{O}_{HNe}+\mbox{h.c.}$ & $\times$ & $2n_f^2$ & $2$ & $18$
\\
\hline
$\psi^2HX+\mbox{h.c.}$
&$\mathcal{O}_{NB}$  & $\times$  &  $2n_f^2$ & $2$ & $18$
\\
& $\mathcal{O}_{NW}$& $\times$  &  $2n_f^2$ & $2$ & $18$
\\
\hline
&$\mathcal{O}_{NN}$& $\mathcal{O}^{prst}_{NN}=\mathcal{O}^{srpt}_{NN}=\mathcal{O}^{ptsr}_{NN}$  & $\frac{1}{4}n_f^2(n_f+1)^2$ & $1$&$36$
\\
& $\mathcal{O}_{eN}$& $\times$ &$n_f^4$& $1$ & $81$
\\
$(\bar{R}R)(\bar{R}R)$
& $\mathcal{O}_{uN}$& $\times$ &$n_f^4$& $1$ & $81$
\\
& $\mathcal{O}_{dN}$& $\times$ &$n_f^4$& $1$ & $81$
\\
& $\mathcal{O}_{duNe}+\mbox{h.c.}$& $\times$ &$2n_f^4$& $2$ & $162$
\\
\hline
$(\bar{L}L)(\bar{R}R)$
&$\mathcal{O}_{LN}$ &$\times$&   $n_f^4$& $1$ & $81$
\\
&$\mathcal{O}_{QN}$ &$\times$&  $n_f^4$& $1$ & $81$
\\
\hline
$(\bar{L}R)(\bar{R}L)+\mbox{h.c.}$
&$\mathcal{O}_{QuNL}$ &$\times$&   $2n_f^4$& $2$ & $162$
\\
\hline
&$\mathcal{O}_{LNLe}$ &$\times$&   $2n_f^4$& $2$ & $162$
\\
$(\bar{L}R)(\bar{L}R)+\mbox{h.c.}$
&$\mathcal{O}_{LNQd}$ &$\times$&  $2n_f^4$& $2$ & $162$
\\
&$\mathcal{O}_{LdQN}$ &$\times$&   $2n_f^4$& $2$ & $162$
\\
\hline
$\slashed{L}\cap B +\mbox{h.c.} $
&$\mathcal{O}_{NNNN}$ &$\mathcal{O}^{prst}_{NNNN}=\mathcal{O}^{stpr}_{NNNN}=\mathcal{O}^{rpst}_{NNNN}$& $\frac{1}{6}n_f^2(n_f^2-1)$ & $0$&$12$
\\
& &$\mathcal{O}^{prst}_{NNNN}=-\mathcal{O}^{pstr}_{NNNN}-\mathcal{O}^{ptrs}_{NNNN}$ & & &
\\
\hline
$\slashed{L}\cap\slashed{B}+\mbox{h.c.}  $
&$\mathcal{O}_{QQdN}$ &$\mathcal{O}_{QQdN}^{prst}-\mathcal{O}_{QQdN}^{rpst}=0$& $n_f^3(n_f+1)$ & $2$ & $108$
\\
&$\mathcal{O}_{uddN}$ &$\times$& $2n_f^4$& $2$ & $162$
\\
\hline
Total with $L\cap B$ & &   & $\frac{1}{4}n_f^2(61n_f^2+2n_f+37) $ & $25$ & $1332 $
\\
\hline
Total with $\slashed{L}\cap B   $ & & & $\frac{1}{6}n_f^2(n_f^2-1)$ & $0$ & $12$
\\
\hline
Total with $\slashed{L}\cap\slashed{B}$  &  & &$ n_f^3(3n_f+1)$ &  $4$ &  $270$
 \\
\hline
 Total &  &  & $\frac{1}{12} n_f^2 (221 n_f^2+18 n_f+109)$ & $ 29$ & $1614 $
\\
\hline
\end{tabular}
\caption{Counting of our dim-6 operators involving $N$ for each independent set of flavors. The symbol $\times$ indicates absence of flavor symmetry relations for relevant operators. Hermitian conjugated operators are included.}
\label{tab2}
\end{table}

The $(19+12=)~31$ operators covering also Hermitian conjugates of the operators listed in Table~\ref{tab1} can also be counted for each independent combination of fermion flavors. To do so, we have to take into account all flavor symmetry relations in the operators, as shown in the third column of Table~\ref{tab2}. The number of relations is then subtracted when counting independent operators, with the end result being given in the last three columns for general $n_f$ generations and for $n_f=1,~3$ in particular. Note that the operator $\mathcal{O}_{NNNN}$ and its conjugate vanish identically when all four $N$s are identical at $n_f=1$. We have also verified this way of counting by working out the Hilbert series by introducing sterile neutrinos into the code of Ref.~\cite{Henning:2015alf}.

\section{Dimension seven operators involving sterile neutrinos}
\label{sec3}

The dim-7 operators involving sterile neutrinos $N$ can be systematically classified according to the number of $N$ fields:
\begin{eqnarray}
&&\{N\psi, N^2\}\otimes\{\varphi^4, \varphi^3D, \varphi^2D^2,  \varphi^2X,\varphi D X,  X^2, \varphi D^3, D^4, D^2X\}
\nonumber
\\
&&\oplus\{N\psi^3,N^2\psi^2, N^3\psi,N^4\}\otimes\{D,\varphi \},
\end{eqnarray}
where $\psi$ can be any SM fermion field, $X$ any gauge field strength, and $\varphi\in\{H, \tilde{H}\}$. Since all odd-dimensional operators carry lepton (and for some also baryon) number~\cite{Degrande:2012wf}, we can choose uniformly all basis operators to have the same sign lepton number. We start with some trivial observations. First, the operator classes in the set $\{N\psi, N^2\}\otimes\{\varphi D^3, D^4, D^2X\}$ can be reduced to others with less covariant derivatives by using IBP, EoM, and Fierz identities. Second, it is easy to check that the classes in the set
$\{N\psi\varphi^4,~N\psi X^2,~N^2\varphi^3D,~N^2\varphi DX,~N^3\psi D,~N^4\varphi\}$ cannot survive the $U(1)_Y\otimes SU(2)_L$ symmetries. We are thus left with the following fourteen classes,
\begin{eqnarray}
\nonumber
&&\{N\psi\varphi^3D, N\psi\varphi^2D^2,  N\psi\varphi^2X, N\psi\varphi D X\}
\oplus\{ N^2\varphi^4, N^2\varphi^2D^2,  N^2 \varphi^2X, N^2X^2\}
\\
&&\oplus\{N\psi^3D,N^2\psi^2D, N^4D\}
\oplus\{N\psi^3\varphi, N^2\psi^2\varphi,  N^3\psi\varphi\},
\end{eqnarray}
which we analyze below one by one.

(1) $N\psi\varphi^3D$ -- The $SU(2)_L\otimes SU(3)_C$ symmetries require $\psi$ to be a doublet lepton field $L$. Taking into account $U(1)_Y$ and IBP, we are left with two independent operators,
\begin{eqnarray}
\nonumber
\mathcal{O}_{NL1}&=&\epsilon_{ij}(NC\gamma_\mu L^i) (iD^\mu H^j)(H^\dagger H),
\\
\mathcal{O}_{NL2}&=&\epsilon_{ij}(NC\gamma_\mu L^i)  H^j
(H^\dagger i\overleftrightarrow{D^\mu} H),
\end{eqnarray}
where $H^\dagger \overleftrightarrow{D}_\mu H=H^\dagger D_\mu H-(D_\mu H)^\dagger H$.

(2) $N\psi\varphi^2D^2$ -- The $SU(2)_L\otimes SU(3)_C$ symmetries require that $\psi$ be a singlet lepton $e$, and then the two $\varphi$s have to be 2 $H$s by $U(1)_Y$ symmetry. We thus have one operator in this class modulo EoM,
\begin{equation}
\mathcal{O}_{NeD}=\epsilon_{ij}(NCD_\mu e)(H^iD^\mu H^j).
\end{equation}

(3) $N\psi\varphi^2X$ -- The same argument as above yields the unique operator,
\begin{equation}
\mathcal{O}_{NeW}=(\epsilon\tau^I)_{ij}(NC\sigma^{\mu\nu}e)(H^iH^j)W^I_{\mu\nu}.
\end{equation}

(4) $N\psi\varphi DX$ -- The gauge symmetries imply $(\psi,\varphi)=(L,H)$. Since the fermion bilinear should be in the form of a four-vector, i.e., $(NC\gamma_\mu L)$, the covariant derivative can only act on the scalar $H$ to avoid the presence of EoM. Considering $X$ can be a gauge field strength tensor or its dual
$\tilde X_{\mu\nu\alpha\beta}=(1/2)\epsilon_{\mu\nu\alpha\beta}X^{\alpha\beta}$, we get four independent operators in this class, 
\begin{eqnarray}\label{nfhdx}
\nonumber
\mathcal{O}_{NLB1}&=&\epsilon_{ij}(NC\gamma^\mu L^i)(D^\nu H^j)B_{\mu\nu},
\\
\nonumber
\mathcal{O}_{NLB2}&=&\epsilon_{ij}(NC\gamma^\mu L^i)(D^\nu H^j)\tilde{B}_{\mu\nu},
\\
\nonumber
\mathcal{O}_{NLW1}&=&(\epsilon \tau^I)_{ij}(NC\gamma^\mu L^i)(D^\nu H^j)W^I_{\mu\nu},
\\
\mathcal{O}_{NLW2}&=&(\epsilon \tau^I)_{ij}(NC\gamma^\mu L^i)(D^\nu H^j)\tilde{W}^I_{\mu\nu}.
\end{eqnarray}

(5) $N^2\varphi^4$ -- This form is uniquely determined to be
\begin{equation}
\mathcal{O}_{NH}=(NCN)(H^\dagger H)^2.
\end{equation}

(6) $N^2\varphi^2D^2$ -- The scalar bilinear can be chosen to be Hermitian. In this case we find two independent operators, while other possible ones can be expressed as a linear combination of them plus EoM operators:
\begin{eqnarray}\nonumber
\mathcal{O}_{ND1}&=&(NCD_\mu N)(H^\dagger \overleftrightarrow{D^\mu} H),
\\
\mathcal{O}_{ND2}&=&(NC N)\Big((D_\mu H)^\dagger D^\mu H\Big).
\end{eqnarray}
The operator $\mathcal{O}_{ND1}$ was missed in Ref.~\cite{Bhattacharya:2015vja}.

(7) $N^2\varphi^2X$ -- In this class the gauge field strength $X$ can be either $B_{\mu\nu}$ or $W^I_{\mu\nu}$, but it is not necessary to consider its dual because of Eq.~(\ref{dirac4}):
\begin{eqnarray}
\nonumber
\mathcal{O}_{NNB}&=&(NC\sigma_{\mu\nu}N)(H^\dagger H)B^{\mu\nu},
\\
\mathcal{O}_{NNW}&=&(NC\sigma_{\mu\nu}N)(H^\dagger
\tau^I H)W^{I\mu\nu},
\end{eqnarray}
where the second one was not included in Ref.~\cite{Bhattacharya:2015vja}.

(8) $N^2 X^2$ -- The operators in this class are found to be consistent with the ones given in Ref.~\cite{Bhattacharya:2015vja}, which are renamed as follows,
\begin{eqnarray}\nonumber
\mathcal{O}_{NB1}&=&(NCN)B_{\mu\nu}B^{\mu\nu},
\\
\nonumber
\mathcal{O}_{NB2}&=&(NCN)B_{\mu\nu}\tilde{B}^{\mu\nu},
\\
\nonumber
\mathcal{O}_{NW1}&=&(NCN)W^I_{\mu\nu}W^{I\mu\nu},
\\
\nonumber
\mathcal{O}_{NW2}&=&(NCN)W^I_{\mu\nu}\tilde{W}^{I\mu\nu},
\\
\nonumber
\mathcal{O}_{NG1}&=&(NCN)G^A_{\mu\nu}G^{A\mu\nu},
\\
\mathcal{O}_{NG2}&=&(NCN)G^A_{\mu\nu}\tilde{G}^{A\mu\nu}.
\end{eqnarray}

(9) $N\psi^3D$ -- The gauge symmetries completely determine the fermion field contents in this class. When IBP and EoM are taken into account, we find six different operators  which can be chosen as,
\begin{eqnarray}\nonumber
\mathcal{O}_{eNLLD}&=&\epsilon_{ij}(\bar{e}\gamma_\mu N)(L^iCiD_\mu L^j),
\\
\nonumber
\mathcal{O}_{duNeD}&=&(\bar{d}\gamma_\mu u)(NCiD_\mu e),
\\
\nonumber
\mathcal{O}_{QLNuD}&=&(\bar{Q}\gamma_\mu L)(NCiD_\mu u),
\\
\nonumber
\mathcal{O}_{dNQLD}&=&\epsilon_{ij}(\bar{d}\gamma_\mu N)(Q^iCiD_\mu L^j),
\\
\nonumber
\mathcal{O}_{dNduD}&=&\epsilon_{\alpha\beta\sigma}(\bar{d}_{\alpha}\gamma_\mu N)(\bar{d}_{\beta}iD_\mu u^C_{\sigma}),
\\
\mathcal{O}_{QdQND}&=&\epsilon_{ij}\epsilon_{\alpha\beta\sigma}(\bar{Q}_{i\alpha}\gamma_\mu d^C_{\beta})(\bar{Q}_{j\sigma}iD_\mu N).
\label{Npsi3D}
\end{eqnarray}
The last two operators violate baryon number by one unit.

(10) $N^2\psi^2D$ -- The same analysis as above yields five independent operators,
\begin{eqnarray}\nonumber
\mathcal{O}_{LND}&=&(\bar{L}\gamma_\mu L)(NCiD^\mu N),
\\
\nonumber
\mathcal{O}_{QND}&=&(\bar{Q}\gamma_\mu Q)(NCiD^\mu N),
\\
\nonumber
\mathcal{O}_{eND}&=&(\bar{e}\gamma_\mu e)(NCiD^\mu N),
\\
\nonumber
\mathcal{O}_{uND}&=&(\bar{u}\gamma_\mu u)(NCiD^\mu N),
\\
\mathcal{O}_{dND}&=&(\bar{d}\gamma_\mu d)(NCiD^\mu N).
\label{N2psi2D}
\end{eqnarray}

(11) $N^4 D$ -- This is easy to figure out:
\begin{equation}
\mathcal{O}_{NND}=(\bar{N}\gamma_\mu N)(NCiD^\mu N).
\label{N4D}
\end{equation}

(12) $N\psi^3\varphi$ -- The gauge symmetries can first determine the field contents, and then Fierz identities are used to transform any four-vector fermion bilinear forms to scalar ones. According to this logic, we find thirteen operators,
\begin{eqnarray}\nonumber
\mathcal{O}_{LNLLH}&=&\epsilon_{ij}(\bar{L}N)(LCL^i)H^j,
\\
\nonumber
\mathcal{O}_{QNQLH1}&=&\epsilon_{ij}(\bar{Q}N)(QCL^i)H^j,
\\
\nonumber
\mathcal{O}_{QNQLH2}&=&\epsilon_{ij}(\bar{Q}N)(Q^iCL^j)H,
\\
\nonumber
\mathcal{O}_{eLNeH}&=&\epsilon_{ij}(\bar{e}L^i)(NCe)H^j,
\\
\nonumber
\mathcal{O}_{dLNdH}&=&\epsilon_{ij}(\bar{d}L^i)(NCd)H^j,
\\
\nonumber
\mathcal{O}_{uLNuH}&=&\epsilon_{ij}(\bar{u}L^i)(NCu)H^j,
\\
\nonumber
\mathcal{O}_{dLNuH}&=&\epsilon_{ij}(\bar{d}L^i)(NCu)\tilde{H}^j,
\\
\nonumber
\mathcal{O}_{dQNeH}&=&\epsilon_{ij}(\bar{d}Q^i)(NCe)H^j,
\\
\nonumber
\mathcal{O}_{QuNeH}&=&(\bar{Q}u)(NCe)H,
\\
\nonumber
\mathcal{O}_{QeNuH}&=&(\bar{Q}e)(NCu)H,
\\
\nonumber
\mathcal{O}_{QNudH}&=&\epsilon_{\alpha\beta\sigma}(\bar{Q}_{\alpha}N)(\bar{u}_\beta d^C_\sigma)H,
\\
\nonumber
\mathcal{O}_{QNddH}&=&\epsilon_{ij}\epsilon_{\alpha\beta\sigma}(\bar{Q}_{i\alpha}N)(\bar{d}_\beta d^C_\sigma)\tilde{H}^j,
\\
\mathcal{O}_{QNQQH}&=&\epsilon_{ij}\epsilon_{\alpha\beta\sigma}(\bar{Q}_{i\alpha}N)(\bar{Q}_{j\beta }Q^C_\sigma)H,
\end{eqnarray}
where the last three violate baryon number by one unit.

(13) $N^2\psi^2\varphi$ -- Similarly,
\begin{eqnarray}
\nonumber
\mathcal{O}_{LNeH}&=&(\bar{L}N)(NCe)H,
\\
\nonumber
\mathcal{O}_{eLNH}&=&H^\dagger(\bar{e}L)(NCN),
\\
\nonumber
\mathcal{O}_{QNdH}&=&(\bar{Q}N)(NCd)H,
\\
\nonumber
\mathcal{O}_{dQNH}&=&H^\dagger(\bar{d}Q)(NCN),
\\
\nonumber
\mathcal{O}_{QNuH}&=&(\bar{Q}N)(NCu)\tilde{H},
\\
\mathcal{O}_{uQNH}&=&\tilde{H}^\dagger(\bar{u}Q)(NCN).
\end{eqnarray}

(14) $N^3\psi\varphi$ -- We have
\begin{eqnarray}
\nonumber
\mathcal{O}_{LNNH}&=&(\bar{L}N)(NCN)\tilde{H},
\\
\mathcal{O}_{NLNH}&=&\tilde{H}^\dagger(\bar{N}L)(NCN).
\end{eqnarray}

In summary, we find there are 52 dim-7 independent operators containing sterile neutrinos. All these operators carry two units of lepton number, five of which further carry one unit of baryon number, and are thus all non-Hermitian. Compared with the counts in Ref.~\cite{Bhattacharya:2015vja}, we find 16 operators in their list are redundant while the list missed 2 new operators. To see better the difference between this work and Ref.~\cite{Bhattacharya:2015vja}, we make detailed comparison in Tables \ref{tab3} and \ref{tab4}. The first two columns in the tables list the classes and operators in \cite{Bhattacharya:2015vja} but with fields named as in current work. For the class $\psi^4D$, we show one, i.e., $(L^\T_1C\sigma^{\mu\nu}L_2)D_\mu(L^\T_3C\gamma_\nu R)$ with $L/R$ referring to left/right-handed fields, of the two equivalent structures given in Ref.~\cite{Bhattacharya:2015vja} without loss of generality. The items $9^2$ and $10^3$ in the second column of Table~\ref{tab4} indicate there are two and three operators respectively according to \cite{Bhattacharya:2015vja}. But we find one operator is redundant in each case, upon using the Schouten identities:
\begin{eqnarray}
\epsilon_{ij}\epsilon_{mn}+\epsilon_{im}\epsilon_{nj}+\epsilon_{in}\epsilon_{jm}=0,~ \delta_{ij}\epsilon_{mn}+\delta_{im}\epsilon_{nj}+\delta_{in}\epsilon_{jm}=0.
\label{SchoutenI}
\end{eqnarray}
Such redundant operators are marked with a $\times$ in the third column of the tables where we show our list of operators. The symbol $=$ ($\sim$) implies the relevant operator in \cite{Bhattacharya:2015vja} is identical with ours (up to a constant), while remaining operators without a prefix are either new ($\mathcal{O}_{ND1}$ and $\mathcal{O}_{NNW}$) or can be made equivalent upon using IBP, EoM, and Fierz identities. We are aware that in most cases the choice of independent operators is not unique. As we stated earlier, our criterion to choose independent operators is to follow as closely as possible the conventions in Refs.~\cite{Grzadkowski:2003tf,Liao:2016hru} and in addition to introduce as few gamma matrices as possible. The latter can be best seen in our choice of operators for the $\psi^4D$ class in Eqs.~(\ref{Npsi3D},\ref{N2psi2D},\ref{N4D}) (involving one gamma matrix) vs their counterparts in the second column of Table~\ref{tab3} (involving three).

As in the case of dim-6 operators we also count the number of dim-7 operators with independent flavor structures in order to compare with the Hilbert series approach. This count is shown for $n_f$ generations of fermions and $n_f=1,~3$ in the last three columns of Tables~\ref{tab3} and \ref{tab4}. Note that a factor of two has to be attached to all numbers when Hermitian conjugates are included. Our counting has been verified also using the code in \cite{Henning:2015alf}.

\begin{table}
\small
\centering
\begin{tabular}{|c|l||l|c|c|c|c|}
\hline
Class in \cite{Bhattacharya:2015vja} &  Operator in \cite{Bhattacharya:2015vja} & This work & $n_f$ & $n_f=1$ & $n_f=3$
\\
\hline
$\psi^2H^4$ & $\overline{N^C}N|H|^4$ & $=\mathcal{O}_{NH}$ & $\frac{1}{2}n_f(n_f+1)$  & 1& 6
\\
\hline
&$(\overline{N^C}\gamma^{\mu}H^\T\epsilon L)(H^\dagger i\overleftrightarrow{D}_{\mu}H)$  & $\sim\mathcal{O}_{NL2}$ & $n_f^2$ & 1&9
\\
$\psi^2H^3D$&$(\overline{N^C}\gamma^{\mu}H^\dagger L)(\tilde{H}^\dagger\overleftrightarrow{D}_{\mu}H)$  & $\mathcal{O}_{NL1}$ & $n_f^2$ & 1&9
\\
&$(\overline{N^C}\gamma^{\mu}H^\T\epsilon L)(\partial_{\mu}|H|^2)$  &  $\times$  & & &
\\
\hline
&$(\overline{N^C}D_{\mu}e)(\tilde{H}^\dagger D^\mu H)$  & $=-\mathcal{O}_{NeD}$ & $n_f^2$  & 1&9
\\
$\psi^2H^2D^2$ &$(\overline{N^C}N)|DH|^2$  &  $=\mathcal{O}_{ND2}$ & $\frac{1}{2}n_f(n_f+1)$ & 1&6
\\
&  &  $\mathcal{O}_{ND1}$ (new!)  & $\frac{1}{2}n_f(n_f-1)$ & 0 &3
\\
\hline
&$(\overline{N^C}\sigma^{\mu\nu}e)(\tilde{H}^\dagger \tau^IW^I_{\mu\nu}H)$  & $=-\mathcal{O}_{NeW}$  & $n_f^2$ & 1&9
\\
$\psi^2H^2X$ &$|H|^2(\overline{N^C}\sigma^{\mu\nu}N)B_{\mu\nu}$  &  $=\mathcal{O}_{NNB}$ & $\frac{1}{2}n_f(n_f-1)$ & 0&3
\\
&  &  $\mathcal{O}_{NNW}$ (new!) & $\frac{1}{2}n_f(n_f-1)$  & 0&3
\\
\hline
&$(\partial^\mu \overline{N^C})\gamma^\nu H^\T\epsilon LB_{\mu\nu} $ &  $\mathcal{O}_{NLB1}$ & $n_f^2$ &1 &9
\\
&$\overline{N^C}\gamma^\mu(\tilde{H}^\dagger D^\nu L)B_{\mu\nu}$   & $\mathcal{O}_{NLB2}$ & $n_f^2$  &1 &9
\\
&$(\partial^\mu \overline{N^C})\gamma^\nu (\tilde{H}^\dagger\tau^IW^I_{\mu\nu}L)$ & $\mathcal{O}_{NLW1}$ & $n_f^2$&1 &9
\\
$\psi^2HDX$ &$\overline{N^C}\gamma^\mu(\tilde{H}^\dagger\tau^IW^I_{\mu\nu} D^\nu L)$   & $\mathcal{O}_{NLW2}$ & $n_f^2$ &1 &9
\\
&$(\partial^\mu \overline{N^C})\gamma^\nu H^\T\epsilon L \tilde{B}_{\mu\nu} $ &  $\times$  & & &
\\
&$\overline{N^C}\gamma^\mu(\tilde{H}^\dagger D^\nu L)\tilde{B}_{\mu\nu}$   &   $\times$  & & &
\\
&$(\partial^\mu \overline{N^C})\gamma^\nu(\tilde{H}^\dagger\tau^I\tilde{W}^I_{\mu\nu}L)$
&  $\times$
& & &
\\
&$\overline{N^C}\gamma^\mu(\tilde{H}^\dagger\tau^I\tilde{W}^I_{\mu\nu} D^\nu L)$   &   $\times$   &  & &
\\
\hline
& $\overline{N^C}N(G^A_{\mu\nu})^2$ & $=\mathcal{O}_{NG1}$ & $\frac{1}{2}n_f(n_f+1)$ &1 &6
\\
& $\overline{N^C}N(W^I_{\mu\nu})^2$ & $=\mathcal{O}_{NW1}$ & $\frac{1}{2}n_f(n_f+1)$ &1 &6
\\
$\psi^2X^2$& $\overline{N^C}N(B_{\mu\nu})^2$ & $=\mathcal{O}_{NB1}$ & $\frac{1}{2}n_f(n_f+1)$ &1 &6
\\
& $\overline{N^C}N(\tilde{G}^A_{\mu\nu}G^A_{\mu\nu})$ & $=\mathcal{O}_{NG2}$ & $\frac{1}{2}n_f(n_f+1)$ &1 &6
\\
& $\overline{N^C}N(\tilde{W}^I_{\mu\nu}W^I_{\mu\nu})$ & $=\mathcal{O}_{NW2}$ & $\frac{1}{2}n_f(n_f+1)$ &1 &6
\\
& $\overline{N^C}N(\tilde{B}_{\mu\nu}B_{\mu\nu})$ & $=\mathcal{O}_{NB2}$ & $\frac{1}{2}n_f(n_f+1)$ &1 &6
\\
\hline
&  $4:(QC\sigma^{\mu\nu}d^C)D_\mu(LC\gamma_\nu N)$ & $\mathcal{O}_{dNQLD}$ & $n_f^4$ & 1 & 81
\\
&  $ 5:(QC\sigma^{\mu\nu}L)D_\mu(d^CC\gamma_\nu N)$ &   $\times$ & & &
\\
&  $6:(d^CC\sigma^{\mu\nu}L)D_\mu(QC\gamma_\nu N)$ &  $\times$ & & &
\\
&  $7:(LC\sigma^{\mu\nu}e^C)D_\mu(LC\gamma_\nu N)$ & $\mathcal{O}_{eNLLD}$  & $\frac{1}{2}n_f^3(n_f+1)$ & 1 & 54
\\
& $8:(QC\sigma^{\mu\nu}u^C)D_\mu(N^CC\gamma_\nu L^C)$ & $\mathcal{O}^\dagger_{QLNuD}$ &  $n_f^4$ & 1 & 81
\\
&  $9:(QC\sigma^{\mu\nu}N^C)D_\mu(u^CC\gamma_\nu  L^C)$ &  $\times$ & & &
\\
& $10:(u^CC\sigma^{\mu\nu} N^C)D_\mu(QC\gamma_\nu  L^C)$ &  $\times$ & & &
\\
& $11:(u^CC\sigma^{\mu\nu}N^C)D_\mu(e^CC\gamma_\nu  d)$ & $\mathcal{O}^\dagger_{duNeD}$ &  $n_f^4$ & 1 & 81
\\
$\psi^4D$& $12:(u^CC\sigma^{\mu\nu}e^C)D_\mu(N^CC\gamma_\nu d)$ &  $\times$ & & &
\\
& $13:(N^CC\sigma^{\mu\nu}e^C)D_\mu(u^CC\gamma_\nu d)$ & $\times$ & & &
\\
& $14:(u^CC\sigma^{\mu\nu}d^C)D_\mu(d^CC\gamma_\nu N)$ & $\mathcal{O}^\dagger_{dNduD}~(\slashed{B})$ & $\frac{1}{2}n_f^3(n_f+1)$ & 1 & 54

\\
& $15:(QC\sigma^{\mu\nu}N^C)D_\mu(QC\gamma_\nu d)$ & $\mathcal{O}^\dagger_{QdQND}~(\slashed{B})$ & $\frac{1}{2}n_f^3(n_f-1)$ & 0 & 27
\\
& $16:(QC\sigma^{\mu\nu} N^C)D_\mu(N^CC\gamma_\nu Q^C)$ &$\mathcal{O}^\dagger_{QND}$  & $\frac{1}{2}n_f^3(n_f-1)$ & 0 & 27
\\
& $17:(u^CC\sigma^{\mu\nu}N^C)D_\mu(N^CC\gamma_\nu   u)$ & $\mathcal{O}^\dagger_{uND}$ & $\frac{1}{2}n_f^3(n_f-1)$ & 0 & 27
\\
& $18:(d^CC\sigma^{\mu\nu}N^C)D_\mu(N^CC\gamma_\nu   d)$ &$\mathcal{O}^\dagger_{dND}$  & $\frac{1}{2}n_f^3(n_f-1)$ & 0 & 27
\\
& $19:(LC\sigma^{\mu\nu}N^C)D_\mu(N^CC\gamma_\nu  L^C)$ &  $\mathcal{O}^\dagger_{LND}$ & $\frac{1}{2}n_f^3(n_f-1)$ & 0 & 27
\\
& $20:(N^CC\sigma^{\mu\nu}e^C)D_\mu(N^CC\gamma_\nu   e)$ & $\mathcal{O}^\dagger_{eND}$  & $\frac{1}{2}n_f^3(n_f-1)$ & 0 & 27
\\
& $21:(N^CC\sigma^{\mu\nu}N^C)D_\mu(N^CC\gamma_\nu N)$ & $\mathcal{O}^\dagger_{NND}$ & $\frac{1}{6}n_f^2(n_f-1)(n_f-2)$ & 0 & 3
\\
\hline
\end{tabular}
\caption{Comparison of dim-7 operators involving sterile neutrinos between Ref.~\cite{Bhattacharya:2015vja} and this work. The numbers in the last three columns are to be multiplied by a factor two when Hermitian conjugated operators are counted.}
\label{tab3}
\end{table}

Now we demonstrate redundancy of operators in \cite{Bhattacharya:2015vja} by a few examples in the classes, $\psi^2H^3D$, $\psi^2HDX$, $\psi^4D$, $\psi^4H$. Consider the three operators in the class $\psi^2H^3D$ of Table~\ref{tab3}. They can be expressed in terms of the two independent operators $\mathcal{O}_{NL1},~\mathcal{O}_{NL2}$ plus others in our basis using IBP, EoM, and Schouten identities:
\begin{eqnarray}
\label{class1}
(\overline{N^C}\gamma^{\mu}H^\T\epsilon L)(H^\dagger i\overleftrightarrow{D}_{\mu}H)&=&-\mathcal{O}_{NL2},
\\
\nonumber
(\overline{N^C}\gamma^{\mu}H^\dagger L)(\tilde{H}^\dagger\overleftrightarrow{D}_{\mu}H)&=&
2i\delta_{ij}\epsilon_{mn}(\overline{N^C}\gamma^{\mu}L^i)H^{*j}(H^m iD_{\mu}H^n)
\\
\nonumber
&\overset{(\ref{SchoutenI})}{=}&
2i\epsilon_{ij}(\overline{N^C}\gamma^{\mu}L^i)(iD_{\mu}H^j)(H^\dagger H)
-2i\epsilon_{ij}(\overline{N^C}\gamma^{\mu}L^i)H^j(H^\dagger iD_{\mu}H)
\\
\nonumber
&\overset{\mbox{IBP}}{=}&3i\epsilon_{ij}(\overline{N^C}\gamma^{\mu}L^i)(iD_{\mu}H^j)(H^\dagger H)+i(\overline{N^C}\gamma^{\mu}L^i)H^j(H^\dagger i\overleftrightarrow{D}_{\mu}H)+\dots
\\
\label{class2}
&=&3i \mathcal{O}_{NL1}+i \mathcal{O}_{NL2}+\dots,
\\
\label{class3}
(\overline{N^C}\gamma^{\mu}H^\T\epsilon L)(\partial_{\mu}|H|^2)&\overset{\mbox{IBP}}{=}&
-i \mathcal{O}_{NL1}+\dots,
\end{eqnarray}
where the dots again stand for the operators obtained by EoM that are already in our basis. We claimed in Table~\ref{tab3} that all operators with a dual field strength in the class $\psi^2HDX$ are redundant. We show this using the operator $\overline{N^C}\gamma^\mu(\tilde{H}^\dagger D^\nu L)\tilde{B}_{\mu\nu}$ as an example. We need the following well-known identities for the $\gamma$-matrix:
\begin{eqnarray}
\label{dirac3}
\gamma^\mu\gamma^\beta\gamma^\nu&=&g^{\mu\beta}\gamma^\nu+g^{\nu\beta}\gamma^\mu-g^{\mu\nu}\gamma^\beta-i\epsilon^{\mu\nu\alpha\beta}\gamma_\alpha\gamma^5,
\\
\label{dirac4}
\sigma_{\mu\nu}P_L(P_R)&=&\frac{i}{2}\epsilon_{\mu\nu\rho\sigma}\sigma^{\rho\sigma}P_L(-P_R).
\end{eqnarray}
The operator is manipulated as follows:
\begin{eqnarray}\label{redun}
\nonumber
2\overline{N^C}\gamma^\mu(\tilde{H}^\dagger D^\nu L)\tilde{B}_{\mu\nu}&\overset{(\ref{dirac1})}{=}&
\tilde{H}^\dagger\Big(\overline{N^C}(\gamma^\mu\slashed{D}\gamma^\nu+\gamma^\mu\gamma^\nu
\slashed{D} )L\Big)\tilde{B}_{\mu\nu}
\\
\nonumber
&\overset{(\ref{dirac3})}{=}&
\tilde{H}^\dagger\Big(\overline{N^C}(-i\epsilon^{\mu\nu\alpha\beta}\gamma_\alpha D_\beta\gamma^5+\gamma^\mu\gamma^\nu\slashed{D} )L\Big)\tilde{B}_{\mu\nu}
\\
\nonumber
&\overset{\mbox{EoM}}{=}&
\frac{i}{2}\epsilon^{\mu\nu\alpha\beta}\epsilon_{\mu\nu\lambda\rho}
\tilde{H}^\dagger(\overline{N^C}\gamma_\alpha D_\beta L)B^{\lambda\rho}
+Y_N(H^\dagger H)(\overline{N^C}\gamma^\mu\gamma^\nu N)\tilde{B}_{\mu\nu}
\\
\nonumber
&\overset{(\ref{dirac2})}{=}&-2i\overline{N^C}\gamma^\mu (\tilde{H}^\dagger D^\nu L)B_{\mu\nu}-\frac{i}{2}Y_N(H^\dagger H)(\overline{N^C}\epsilon_{\mu\nu\lambda\rho}\sigma^{\mu\nu}N)B^{\lambda\rho}
\\
&\overset{(\ref{dirac4})}{=}&-2i\overline{N^C}\gamma^\mu (\tilde{H}^\dagger D^\nu L)B_{\mu\nu}+Y_N(H^\dagger H)(\overline{N^C}\sigma^{\mu\nu}N)B_{\mu\nu}.
\end{eqnarray}
The first term in the last equality is indeed the second operator shown in the $\psi^2HDX$ class of Table~\ref{tab3}, while the second term is the second operator in the $\psi^2H^2X$ class (equal to our $\mathcal{O}_{NNB}$) multiplied by the Yukawa matrix $Y_N$. Once the redundancy in the $\psi^2HDX$ class is established, we choose our four independent operators from a different consideration as shown in Eq.~(\ref{nfhdx}).

For the class $\psi^4D$, we establish some equivalence relations which will make redundancy in Ref.~\cite{Bhattacharya:2015vja} evident. As mentioned earlier, the operators in this class are cast \cite{Bhattacharya:2015vja} in one of the equivalent forms, $(L^\T_1C\sigma^{\mu\nu}L_2)D_\mu(L^\T_3C\gamma_\nu R)$, which may be transformed as follows:
\begin{eqnarray}
\nonumber
(L^\T_1C\sigma^{\mu\nu}L_2)D_\mu(L^\T_3C\gamma_\nu R)&\overset{(\ref{dirac2})}{=}&
-i(L^\T_1C\gamma^\nu\gamma^\mu L_2)D_\mu(L^\T_3C\gamma_\nu R)
+i(L^\T_1CL_2)D^\mu(L^\T_3C\gamma_\mu R)
\\
\nonumber
&\overset{\mbox{IBP}}{=}&i(L^\T_2C\slashed{D}\gamma^\mu L_1)(L^\T_3C\gamma_\mu R)+i(L^\T_1C\gamma^\mu\slashed{D} L_2)(L^\T_3C\gamma_\mu R)+\dots
\\
\nonumber
&\overset{(\ref{dirac1})}{=}&2(L^\T_2CiD^\mu L_1)(L^\T_3C\gamma_\mu R)
-i(L^\T_2C\gamma^\mu\slashed{D} L_1)(L^\T_3C\gamma_\mu R)+\dots
\\\label{relation1}
&=&2(L^\T_3C\gamma_\mu R)(L^\T_2CiD^\mu L_1)+\dots
\\
\nonumber
&\overset{(\ref{fierz4})}{=}&-2(L^\T_2C\gamma_\mu R)(L^\T_3CiD^\mu L_1)+(R^\T Ci\slashed{D}L_1)(L^\T_3C L_2)+\dots
\\\label{relation2}
&=&-2(L^\T_2C\gamma_\mu R)(L^\T_3CiD^\mu L_1)+\dots
\\\label{relation3}
&\overset{\mbox{IBP}}{=}&2(L^\T_2C\gamma_\mu R)(L^\T_1CiD^\mu L_3)+\dots,
\end{eqnarray}
where again the dots stand for the operators obtained through EoM that are already covered in the basis. Because of Eqs.~(\ref{relation1},\ref{relation2},\ref{relation3}), we have the following equivalence sequence,
\begin{equation}
\label{relation}
-2(L^\T_1C\gamma_\mu R)(L^\T_2CiD^\mu L_3) \sim (L^\T_1C\sigma^{\mu\nu}L_2)D_\mu(L^\T_3C\gamma_\nu R)\sim (L^\T_2C\sigma^{\mu\nu}L_3)D_\mu(L^\T_1C\gamma_\nu R)\sim (L^\T_3C\sigma^{\mu\nu}L_1)D_\mu(L^\T_2C\gamma_\nu R).
\end{equation}
From the above equation we conclude there is only one independent structure among the three possible ones that associate $R$ with one of $L$s. Considering this we find six redundant operators in this class as shown in Table~\ref{tab3}.

Finally there are five redundant operators in the class $\psi^4H$, as they can be transformed into the chosen ones with the same field contents. We take the operator $14:~(LCe^C)(N^CCN^C)\tilde{H}$ in Table~\ref{tab4} as an example. We attach the flavor indices $p,~r$ to $N$ for better understanding \footnote{Some $SU(2)_L$ contractions were incompletely or incorrectly done in \cite{Bhattacharya:2015vja}. We leave such operators untouched in Table~\ref{tab4} but write explicitly the contraction in Eq.~(\ref{redunx})}:
\begin{eqnarray}
\nonumber
\delta_{ij}H^\dagger_i(L_jCe^C)(N^C_pCN^C_r)&=&\delta_{ij}H^\dagger_i(\bar{e}L_j)(\bar{N}_pN^C_r)
\\
\nonumber
&\overset{(\ref{fierz3})}{=}&
-\delta_{ij}H^\dagger_i(\bar{e}N^C_p)(\overline{L^C_j}N^C_r)
-\delta_{ij}H^\dagger_i(\bar{e}N^C_r)(\overline{L^C_j}N^C_p)
\\
&=&-\delta_{ij}H^\dagger_i(L_jCN^C_r)(N^C_pCe^C)
-\delta_{ij}H^\dagger_i(L_jCN^C_p)(N^C_rCe^C),
\label{redunx}
\end{eqnarray}
where the two operators on the right correspond to the operator $15:~(LCN^C)(N^CCe^C)\tilde{H}$, or $\mathcal{O}^\dagger_{LNeH}$ in our convention with the flavor indices $p,~r$ interchanged.

\begin{table}
\small
\centering
\begin{tabular}{|c|l||l|c|c|c|}
\hline
Class in \cite{Bhattacharya:2015vja} &  Operator in \cite{Bhattacharya:2015vja} & This work & $n_f$ & $n_f=1$ & $n_f=3$
\\
\hline
& 
type $(L^\T_1CL_2)(L^\T_3C L_4)\varphi$:     & & & &
\\
&  $7:(QCu^C)(N^CCe^C)\tilde{H}$ & $\sim\mathcal{O}^\dagger_{QuNeH}$  & $n_f^4$ & 1 & 81
\\
&  $8:(QCe^C)(N^CCu^C)\tilde{H}$ & $\sim\mathcal{O}^\dagger_{QeNuH}$ & $n_f^4$ & 1 & 81
\\
&  $9^2:(QCQ)(QCN^C)\tilde{H}$ & $\times$, $\sim\mathcal{O}^\dagger_{QNQQH}~(\slashed{B})$ & $\frac{1}{3}n_f^2(2n_f^2+1)$ & 1 & 57
\\
&  $10:(QCu^C)(N^CCN^C)H$ &   $\times$ & & &
\\
&  $11:(QCd^C)(N^CCN^C)\tilde{H}$ &  $\times$  &  & &
\\
&  $12:(QCN^C)(N^CCu^C)H$ & $\sim\mathcal{O}^\dagger_{QNuH}$ & $n_f^4$ & 1 & 81
\\
&  $13:(QCN^C)(N^CCd^C)\tilde{H}$ &$\sim\mathcal{O}^\dagger_{QNdH}$  & $n_f^4$ & 1 & 81
\\
&  $ 14:(LCe^C)(N^CCN^C)\tilde{H}$ &  $\times$  &  &  &
\\
&  $15:(LCN^C)(N^C C e^C)\tilde{H}$ & $\sim\mathcal{O}^\dagger_{LNeH}$ & $n_f^4$ & 1 & 81
\\
&  $16:(LCN^C)(N^C C N^C)H$ &  $\sim\mathcal{O}^\dagger_{LNNH}$ & $\frac{1}{3}n_f^2(n_f^2-1)$ & 0 & 24
\\
$\psi^4H$& 
type $(L^\T_1CL_2)(R^\T_1C R_2)\varphi$:            &  &  &  &
\\
&  $5:(QCd^C)(N Ce)H$ & $\sim\mathcal{O}_{dQNeH}$ &  $n_f^4$ & 1 & 81
\\
&  $6:(u^CCL)(uCN)H$ & $\sim\mathcal{O}_{uLNuH}$ & $n_f^4$ & 1 & 81
\\
&  $7:(d^CCL)(uCN)\tilde{H}$ & $\sim\mathcal{O}_{dLNuH}$ & $n_f^4$ & 1 & 81
\\
&  $8:(d^CCL)(dCN)H$ & $\sim\mathcal{O}_{dLNdH}$ & $n_f^4$ & 1 & 81
\\
&  $9:(LCe^C)(N Ce)H$ & $\sim\mathcal{O}_{eLNeH}$  & $n_f^4$ & 1 & 81
\\
&  $10^3:(QCL)(Q^CCN)H$ & $\times, \sim\mathcal{O}_{QNQLH1},\mathcal{O}_{QNQLH2}$ & $2n_f^4$ & 2 & 162
\\
&  $11:(LCL)(L^CCN)H$ & $\sim\mathcal{O}_{LNLLH}$ &  $n_f^4$ & 1 & 81
\\
&  $12:(QCN^C)(uCd)\tilde{H}$ & $\sim\mathcal{O}^\dagger_{QNudH}~(\slashed{B})$ & $n_f^4$ & 1 & 81
\\
&  $13:(QCN^C)(d C d)H$ & $\sim\mathcal{O}^\dagger_{QNddH}~(\slashed{B})$ & $\frac{1}{2}n_f^3(n_f-1)$ & 0 & 27
\\
&  $14:(QCu^C)(N CN)H$ & $\sim\mathcal{O}_{uQNH}$ & $\frac{1}{2}n_f^3(n_f+1)$ & 1 & 54
\\
&  $15:(QCd^C)(N CN)\tilde{H}$ & $\sim\mathcal{O}_{dQNH}$ & $\frac{1}{2}n_f^3(n_f+1)$ & 1 & 54
\\
&  $16:(LCe^C)(N CN)\tilde{H}$ & $\sim\mathcal{O}_{eLNH}$ & $\frac{1}{2}n_f^3(n_f+1)$ & 1 & 54
\\
&  $17:(LCN^C)(N CN)H$ & $\sim\mathcal{O}_{NLNH}$ & $\frac{1}{2}n_f^3(n_f+1)$ & 1 & 54
\\
\hline
Total with $\slashed{L}\cap B$ & &
& $\frac{1}{2}n_f(43n_f^3-n_f^2+27n_f+5)$ & 37 & 1857
\\
\hline
Total with $\slashed{L}\cap\slashed{B}$ &  &  &   $\frac{1}{6}n_f^2(19n_f^2-3n_f+2)$ & 3 & 246
\\
\hline
Total &  &  & $\frac{1}{6}n_f(148n_f^3-6n_f^2+83n_f+15)$ & 40 &2103
\\
\hline
\end{tabular}
\caption{Continuation of comparison between Ref.~\cite{Bhattacharya:2015vja} and this work.}
\label{tab4}
\end{table}

\section{Conclusion}
\label{sec4}

We have made a systematical analysis on the higher dimensional operators up to mass dimension seven in the standard model effective field theory extended with sterile neutrinos. Our study was based on extensive applications of integration by parts, equations of motion, and various Fierz and group identities. We determined the complete and independent set of operators that involve sterile neutrinos, and found that both dimension-six and -seven operators in the previous literature were redundant while two dimension-seven operators were missed. We also counted our operators according to their flavor structures upon taking into account their flavor symmetries, and verified our counting by the Hilbert series approach.

\vspace{0.5cm}
\noindent %
\section*{Acknowledgement}

This work was supported in part by the Grants No. NSFC-11025525, No. NSFC-11575089 and by the CAS Center for Excellence in Particle Physics (CCEPP). Our Hilbert series analysis was based on a slight modification to the Mathematica code provided in Ref.~\cite{Henning:2015alf}.

\noindent %

\end{document}